 \documentclass[traditabstract]{aa}                                  % (traditional abstract) 
\usepackage{txfonts}
\usepackage{natbib}
\bibpunct{(}{)}{,}{a}{}{,}

\usepackage{graphicx}% Include figure files
\usepackage{dcolumn}% Align table columns on decimal point
\usepackage{bm}% bold math
\usepackage{ulem}

\def\be{\begin{equation}}
\def\ee{\end{equation}}
\def\beqn{\begin{eqnarray}}
\def\eeqn{\end{eqnarray}}
\def\eg{E_{\gamma}}

\begin{document}

\title{Introducing the Dirac-Milne universe}

\author{A. Benoit-L{\'e}vy
	\inst{1}
	\and
	G. Chardin\inst{2}
}

\institute{UPMC-CNRS, UMR7095, Institut d'Astrophysique de Paris, F-75014, Paris, France.
\email{benoitl@iap.fr}
\and 
CSNSM, Universit{\'e} Paris-Sud, CNRS-IN2P3, 91405 Orsay, France.
\email{chardin@csnsm.in2p3.fr}
}
\date{\today}

\abstract{The $\Lambda$CDM standard model, although an excellent parametrization of the present cosmological data, contains two as yet unobserved components, dark matter and dark energy, that constitute more than 95\% of the Universe. Faced with this unsatisfactory situation, we study an unconventional cosmology, the Dirac-Milne universe, a matter-antimatter symmetric cosmology, in which antimatter is supposed to present a negative active gravitational mass.  The main feature of this cosmology is the linear evolution of the scale factor with time, which directly solves the age and horizon problems of a matter-dominated universe. We study the concordance of this model to the cosmological test of type Ia supernov\ae\ distance measurements  and calculate the theoretical primordial abundances of light elements for this cosmology. We also show  that the acoustic scale of the cosmic microwave background naturally emerges at the degree scale despite an open geometry.

}
\maketitle

\section{Introduction}

The inflation-based $\Lambda$CDM cosmological model represents a great achievement in modern cosmology. Its predictions are in good agreement with a wide varieties of observational cosmological tests, ranging from big-bang nucleosynthesis and cosmic microwave background anisotropies to distances measurements using type Ia supernov\ae.
However, this success comes at an expensive cost as two poorly-understood components are required in the theory to provide the concordance of the cosmological model to the observations. These components are dark matter, whose density is estimated to be around $\sim 20\%$ of the critical density, and dark energy, which accounts for as much as $\sim75\%$ of the energy content of the Universe.

The physical nature of dark energy remains to be elucidated but all the cosmological observations tend to indicate that it behaves in a similar way as a vacuum energy, characterized by an equation of state $p=-\rho$, where pressure and density are equal in absolute value but opposite. Dark energy  therefore appears to be responsible for the acceleration of the expansion of the Universe, which tends to be confirmed notably by type Ia supernov\ae\ distance measurements.

Faced with this uncomfortable situation where we understand less than 5\% of our Universe, we study the unconventional cosmology of a symmetric universe {\it i.e.} containing equal quantities of matter and antimatter, where antimatter is supposed to present a negative active gravitational mass. The main consequence of this provoking hypothesis is that on large scales,  above the characteristic length of the matter-antimatter distribution, the expansion factor evolves linearly with time, which is reminiscent of the Milne cosmology \citep{Milne1933}.

The main motivation for considering the possibility that antimatter is endowed with a negative active gravitational mass is provided by the study of \cite{Carter68} of the Kerr-Newman geometry representing charged spinning black holes. As noted initially by Carter, the Kerr-Newman geometry with the mass, charge, and spin of an elementary particle such as an electron bears several of the features expected of the real corresponding particle. In particular, the Kerr-Newman ``electron'' has no horizon, has automatically the $g=2$ gyromagnetic ratio, and has a ring structure of radius equal to half the Compton radius of the electron. 
In addition,  this geometry has charge and mass/energy reversal symmetries that strongly evoke the CP and T matter-antimatter symmetry \citep{Chardin97, chardin_rax}: when the non-singular interior of the ring is crossed, a second $\mathbb{R}^4$ space is found where charge and mass change sign \citep{Carter68}. Therefore, starting from an electron of negative charge and positive mass as measured in the first $\mathbb{R}^4$ space, we find in the second space a ``positron" of positive charge and negative mass. The relation of the Kerr-Newman geometry to Dirac's equation  and therefore to antimatter has been noted by some authors \citep{Arcos_04, Burinskii08}.

The introduction of negative  masses is not a popular idea amongst physicists and here we do not attempt to demonstrate that antimatter has a negative active gravitational mass. We  consider this question instead from a cosmological point of view, by studying the properties of this alternative cosmology and deriving necessary conditions that this model must fulfill to comply with observational tests.  Nevertheless, it is important to recall that an analogous system exists in condensed matter, where electrons and holes in a semiconductor appear as quasiparticles with positive inertial mass, but where holes have an energy density lower than the ``vacuum'' constituted by the semiconductor in its ground state and antigravitate in a gravitational field \citep{Tsidilkovski75}.

The gravitational behavior of antimatter has yet to be investigated in laboratory experiments. Gravity experiments on antihydrogen are under way \citep{Aegis2007, Perez2008} and are expected to provide important insight into this behavior, hopefully within the next decade.

The paper is organized as follows. In Sects. \ref{sec:DM} and \ref{sec:therm}, we present some general properties of the Dirac-Milne universe. Section \ref{sec_bbn} is devoted to the study of primordial nucleosynthesis within the framework of the Dirac-Milne cosmology. In \mbox{Sect. \ref{snia}}, we study the compliance of the model to type Ia supernov\ae\ distance measurements.  In Sect. \ref{CMB}, we investigate the geometric implications {\bf on} CMB anisotropies and baryonic acoustic oscillations.

\section{The Dirac-Milne universe\label{sec:DM}}

We consider a model in which space-time is endowed with a Friedmann-Robertson-Walker (FRW) metric, given by the formula

%\begin{equation}
%ds^2=dt^2-a(t)^2\left(\frac{dr^2}{1-k r^2}+r^2d\Omega^2\right),
%\end{equation} 

\begin{equation}
ds^2=dt^2-a(t)^2\left(d\chi^2+\rm{sink}^2 \chi d\Omega^2\right),
\label{metric1}
\end{equation} 
where $\rm{sink}=\sin \chi, \chi, \sinh \chi$ according to the value $k=1, 0, -1$ of the parameter $k$ describing the spatial curvature,  and $a(t)$ is the scale factor.
For such a FRW metric, the time-time and space-space components of the Einstein tensor $G_{\mu\nu}$ take the forms

\be
G_0^0=3 \frac{\dot{a}^2+k}{a^2}   \qquad G_i^i= -\frac{k+2\ddot{a}a+\dot{a}^2}{a^2}.
\label{eq1}
\ee
On scales larger than the characteristic size of the emulsion, the presence of equal quantities of matter with positive mass and antimatter with negative mass nullifies the stress-energy tensor $T_{\mu\nu}$.

Using Einstein equation $G_{\mu\nu}= 8\pi G T_{\mu\nu}$, we therefore have the equivalence
\be
T_{\mu\nu}=0 \Leftrightarrow a(t)\propto t \;\rm{ and }\;  k=-1.
\ee
The metric of the Dirac-Milne universe then reads
\begin{equation}
ds^2=dt^2-t^2\left(d\chi^2+\sinh^2\chi d\Omega^2\right).
\label{metric2}
\end{equation} 

As in the standard $\Lambda$CDM cosmology, the Dirac-Milne universe has a distinctive geometry: while the standard $\Lambda$CDM model has a curved space-time and flat spatial sections, the Dirac-Milne universe has a flat space-time and negatively curved spatial sections.

The Friedmann equation for the Dirac-Milne universe reads
\be
H^2=H_0^2\left(\frac{a_0}{a}\right)^2,
\label{fried1}
\ee
where $a_0$ is the present value of the scale factor. Integrating  this equation enables one to compute the present age of the Universe $t_U$ 
\be
t_U=\frac{1}{H_0},
\ee
where $H_0$ is the Hubble constant. It should be emphasized that this is a strict equality, whereas in the standard $\Lambda$CDM model, the age of the Universe is only approximatively equal to $1/H_0$. It should also be noted that the linear evolution of the scale factor solves the age problem of the Universe \citep{Chaboyer98}, which was a prime concern before the introduction of dark energy,  and does not  affect the  Dirac-Milne  cosmology.

Using the metric in Eq. (\ref{metric1}), the particle horizon, {\it i.e.}, the distance a photon can travel since the origin, is given by the limit
\be
\lim_{t_0\rightarrow 0} \quad \int^{t_U}_{t_0}\frac{dt^\prime}{a(t^\prime)},
\ee
which diverges logarithmically in the case of the Dirac-Milne universe. This simple relation has profound implications as it signifies that any two given places in space were causally connected in the past. The Dirac-Milne universe therefore does not have an horizon problem, which removes the main motivation for the introduction of inflation theories.

Therefore, the linear evolution of the scale factor and the property that the space-time is flat and spatial sections are open naturally solves two major problems in standard cosmology without requiring additional ingredients such as dark energy or inflation, and is  an important motivation for studying in detail such a cosmology.
 
Before presenting the properties of the Dirac-Milne universe and for the sake of clarity, we list the underlying hypothesis and necessary constraints that have been assumed so far:

\begin{enumerate}

\item  The existence of an efficient mechanism for matter and antimatter separation. 
We also assume that the Dirac-Milne universe is composed of separated domains of matter and antimatter.  These universes have been abundantly studied in the context of inhomogeneous nucleosynthesis. Since this universe is matter-antimatter symmetric, this system is a percolation emulsion, meaning that the probability that a given domain extends to infinity is close to unity. It remains unclear, however, whether there is an efficient mechanism to produce such an emulsion. A fundamental parameter is the typical size of this emulsion. We define it as the ratio of the volume $V$ to the surface $S$ separating the two phases: $L=V/S$.  At this stage of work, this characteristic size is a free parameter,  which is assumed here to be a constant, that is constrained using primordial nucleosynthesis (cf Sect. \ref{sec_bbn}).\\

\item Antimatter has a negative active gravitational mass and there is a gravitational repulsion between matter and antimatter. This hypothesis is a necessity to avoid contact between matter and antimatter after cosmological recombination. Indeed, it has been demonstrated \citep{Ramani1976,Cohen1998} that in the case of a symmetric universe, where antimatter is assumed to have a positive gravitational mass, the global size of domains of matter and antimatter must  be  approximatively that of the observable universe. Otherwise, annihilations at the frontiers of matter and antimatter domains would generate a diffuse gamma ray emission that would be in contradiction with observational data.\\

\item  One of the main interests of the Dirac-Milne universe consists in providing a solution to the horizon problem without resorting to inflation. However, this in turn relies on the hypothesis that the contribution of radiation to the stress-energy tensor is negligible. If we attribute to radiation a positive contribution to the energy, the evolution of the universe will not differ significantly from the standard evolution for all epochs where the universe is radiation dominated, i.e. for redshifts above a few thousand. However, in the presence of equal quantities of particles with positive and negative masses, this might no longer be the case. A first step toward a full demonstration that the stress-energy tensor of radiation averages to zero can be deduced from the papers of \citet{Hoyle64}, in the case studied by \citet{Hawking65}, where the particle masses are of both signs and in equal numbers. In this case, it can be shown that the stress-energy tensor of radiation is of order $1/N$ compared to its value when all masses are of the same sign, and where $N$ is the number of mass carriers (particles) in a Hubble horizon.

\end{enumerate}

\section{Some aspects of the thermal history of the Dirac-Milne Universe\label{sec:therm}}
Some simple properties of the Dirac-Milne universe are common to those of a purely linear cosmology studied in \citet{Simmering98}, \citet{Kaplinghat99, Kaplinghat00}, and \citet{Sethi05}. We briefly recall some of these properties.

\subsection{Time-temperature relation}
Using Eq. (\ref{fried1}), it is straightforward to obtain the relation between the age of the Universe and the redshift, hence the temperature
\be
t=\frac{1}{H_0}\frac{T_0}{T},
\ee
where $T_0$ is the present temperature of the Universe, as measured by CMB experiments, $T_0=2.725\pm 0.001\;\rm{K}$ \citep{Fixsen2002}.
\begin{figure}
\begin{center}
\includegraphics[width=\columnwidth]{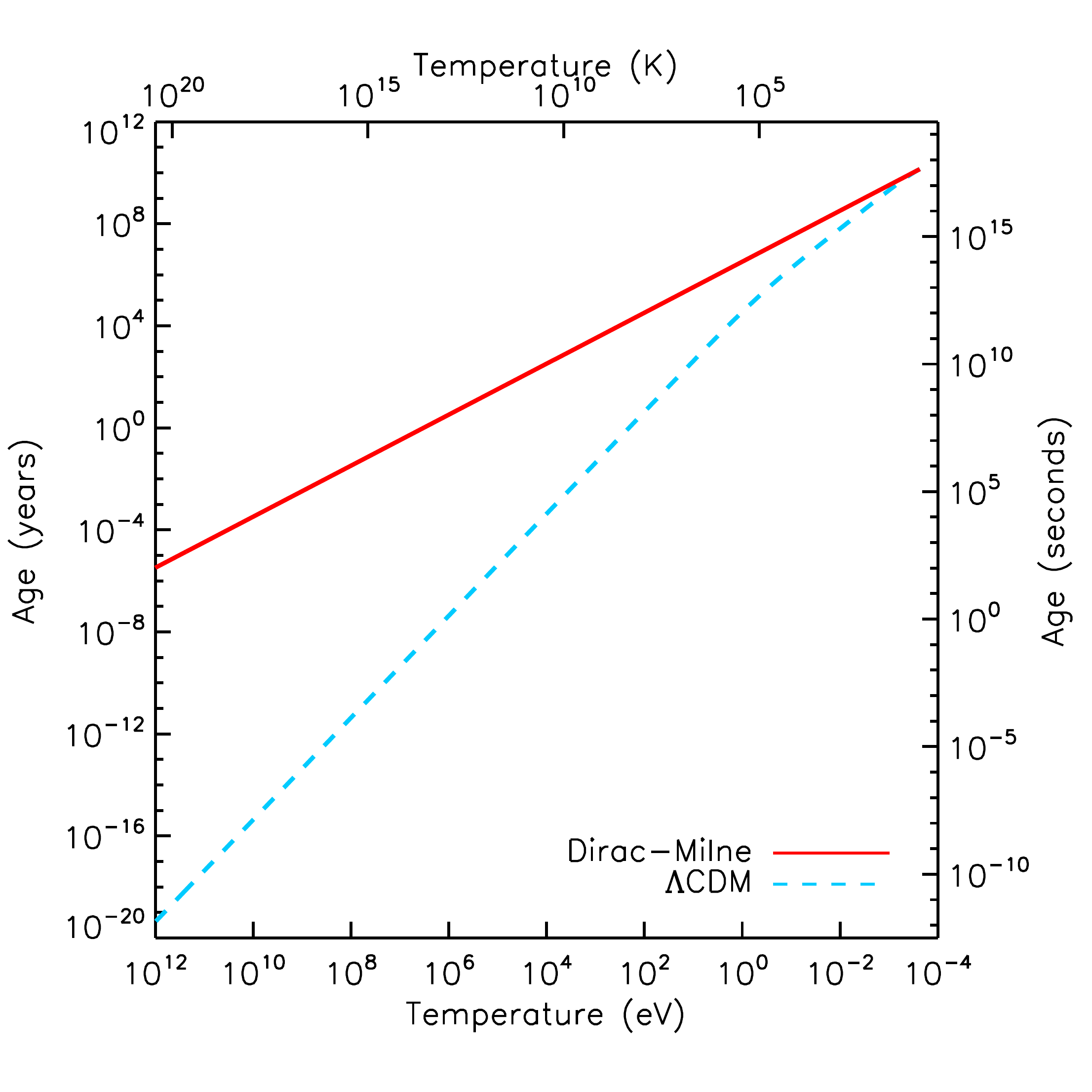}
\caption{\label{age} Age of the universe  as a function of temperature for the Dirac-Milne universe (full line, red) and for the fiducial $\Lambda$CDM cosmology (dashed line, blue).}
\end{center}
\end{figure}
This relation between time and temperature is valid throughout the whole history of the universe and implies that the thermal history of the Dirac-Milne universe is drastically modified from the evolution in the standard $\Lambda$CDM cosmology. Figure (\ref{age}) represents the age of the Universe as a function of the temperature for the Dirac-Milne and the $\Lambda$CDM models. It can be seen that, at high temperatures, the Dirac-Milne universe is much older than the corresponding $\Lambda$CDM cosmology. For instance, the traditional 1 MeV$\sim$1 sec approximation for the standard model becomes 1  MeV $\sim$ 3.3 years in the Dirac-Milne cosmology. As noted in \citet{Simmering98}, this difference has profound implications for big-bang nucleosynthesis calculations, and is discussed in section \ref{sec_bbn}. 

Another temperature of interest is the temperature of the quark-gluon-plasma (QGP) transition. \citet{Omnes1972} proposed that matter-antimatter separation occurred around that temperature, owing to a putative repulsive interaction between nucleons and antinucleons. The maximum size of a domain of (anti)matter was controlled by the diffusion of neutrons. \citet{Aly_sep1974} found a maximum size of $7\times 10^{-4}\;\rm{cm}$ at a temperature of $T\sim330 $ MeV, which was at this epoch the estimated temperature of the QGP transition. This size was later found to differ from the minimum size a domain should have in order to ensure a production of primordial helium compatible with observations \citep{Combes75,Aly1978}.

The situation is rather different in the Dirac-Milne universe as the timescale of the QGP transition is much longer. Since the temperature of the transition is estimated today  to be around $T \sim 170$ MeV \citep{Schwarz2003}, it corresponds to an age of $6\times 10^{5}\;\rm{sec}$ in the Dirac-Milne universe, which is a factor $\sim 10^{10}$ older than in the standard case. This implies that the maximum size to which a domain could possibly grow (assuming the existence of an efficient separation mechanism) is five orders of magnitude higher. For this reason, the Dirac-Milne universe  is far more weakly  constrained by observations than the Omn{\`e}s cosmology.

\subsection{Weak interaction decoupling}

A fundamental example of the modifications induced by a linear scale factor can be seen in  the epoch of decoupling of the weak interactions. This example was analyzed extensively in \citet{Simmering98}, so we only provide here the main results.\begin{figure}
\begin{center}
\includegraphics[width=\columnwidth]{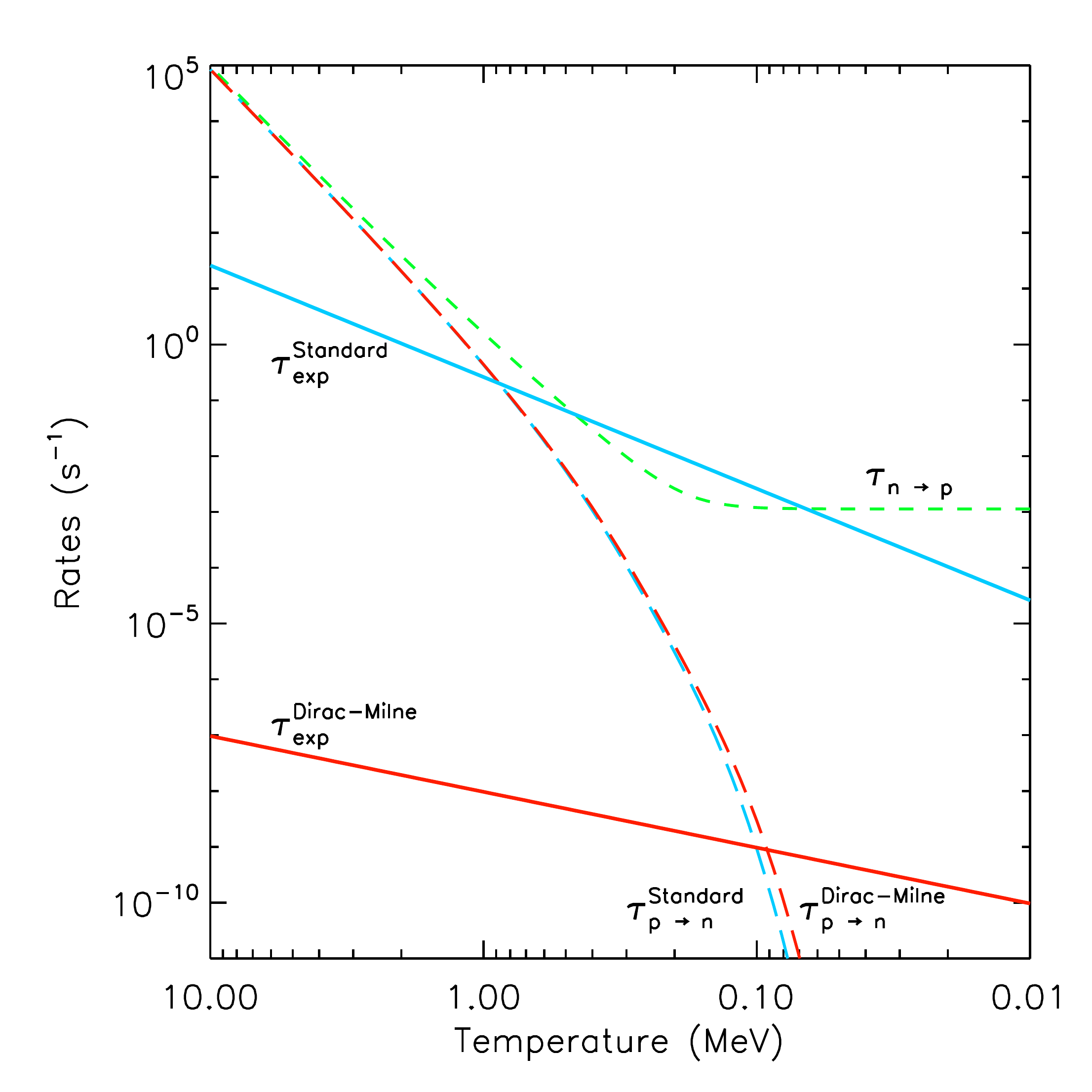}
\caption{\label{wkrates} Comparison between expansion rates and weak reaction rates. The short-dashed green line represents the neutron-to-proton conversion rate. The two solid lines represent the expansion rate of the Dirac-Milne universe (red) and the standard cosmology (blue). The long dashed lines represent the proton-to-neutron conversion rates. Weak interactions decouple when these rates become lower than the expansion rate.}
\end{center}
\end{figure}
In the standard model, weak decoupling occurs at a temperature $T\sim 1\;\rm{MeV}$. In the Dirac-Milne universe, this decoupling happens at a lower temperature of around $T\sim 80 \;\rm{keV}$,  because of the slower variation and the lower value of the expansion rate. This effect is illustrated in Fig. \ref{wkrates}. Weak interactions  control the $n \leftrightarrow p $ equilibrium. At low temperatures, this reaction is limited to the free neutron disintegration (green short-dashed line), but at temperatures higher than 80 keV, the equilibrium between proton and neutrons remains possible. The long-dashed line represent the proton conversion rate as a function of the temperature for the Dirac-Milne universe (red) and the standard cosmology (blue). The analytical expressions for these reaction rates come from \citet{Wagoner69} and \citet{Dicus82}. Weak interactions decouple when the expansion rate  becomes higher than the  $p \leftrightarrow n$ rate. The small difference in the  $p \leftrightarrow n$ rate between the two cosmologies is caused by a difference in the neutrino background temperature. As the weak interactions decouple at a temperature of $T\sim 80\;\rm{keV}$, neutrinos indeed also decouple from the photon background also at this temperature, but only after the annihilation of most of the electron-positron pairs. This implies, as noted in \citet{Simmering98}, that the cosmic neutrino background should have a temperature equal to that of the CMB. This constitutes a distinctive feature of the Dirac-Milne universe.

\section{\label{sec_bbn}Primordial nucleosynthesis}

The primordial nucleosynthesis is a key success of the standard model of cosmology. Theoretical predictions and observations are in good agreement for $^{4}$He, $^{3}$He, and D abundances, although some discrepancies exist for $^{7}$Li (see \citet{Steigman07, Cyburt08} for reviews of standard big-bang nucleosynthesis (SBBN)).
Given the tremendous change in the timescale (of a factor $\sim 10^8$ at $T=1 \;\rm{MeV}$), important modifications to the traditional BBN mechanism arise \citep{Simmering98, Kaplinghat99, Kaplinghat00}. Primordial nucleosynthesis in the Dirac-Milne universe can be described as a two-step process: first the thermal and homogeneous production of $^4$He and $^7$Li, and second, the production of D and $^3$He, the latter being  one of the main novelties introduced by the matter-antimatter symmetry.

\subsection{Thermal and homogeneous BBN}

The age of the universe at a temperature of $T\sim 80 \;\rm{keV}$, below which deuterium is able to survive its photodisintegration, allowing the formation of $^4$He nuclei, is about 40 years in the Dirac-Milne universe. This extremely long timescale, relative to the mean lifetime of the neutron, was initially considered as an impossible obstacle to any kind of primordial helium production in a linear cosmology \citep{Kaplinghat99}. However, it has been pointed out  \citep{Simmering98, Kaplinghat00} that  $^4$He production is possible but relies on a somewhat different mechanism than the standard one. 

As discussed in the previous sections, weak interactions decouple in the Dirac-Milne universe at the low temperature of $T\sim 80\;\rm{keV}$. This implies that the neutron and proton populations remain in thermal equilibrium until that temperature. The ratio of their densities is therefore regulated by the Boltzmann factor
\begin{equation}
\frac{n}{p}=\exp\left(-\frac{Q}{T}\right),
\end{equation}
where $Q=1.29\;\rm{MeV}$ is the difference between the neutron and proton mass.

\begin{figure}
\includegraphics[width=\columnwidth]{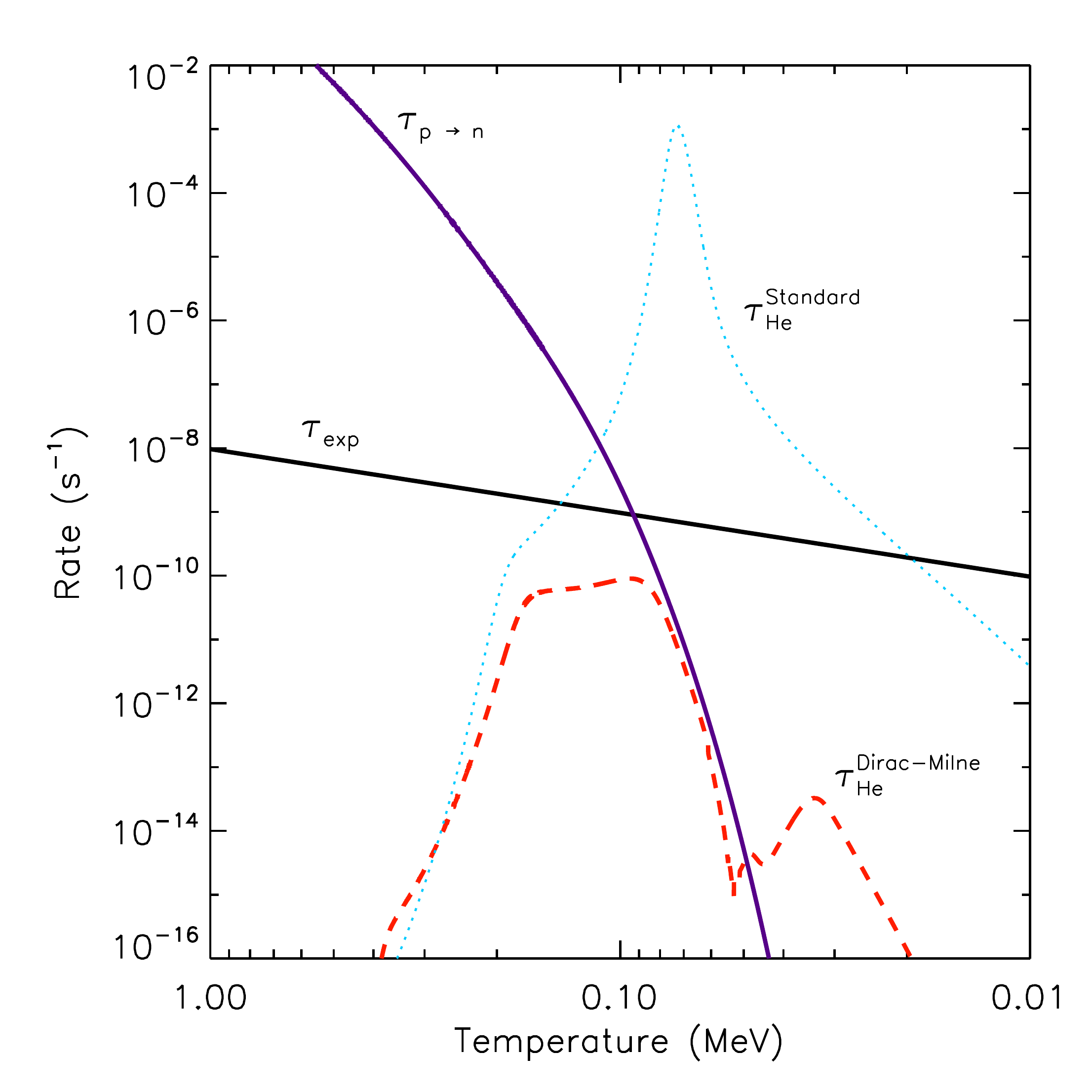}
\caption{\label{dhedt}$^4$He production rate for the Dirac-Milne (dashed red line). The rate drops abruptly when weak interactions decouple around \mbox{$T\sim 80\;\rm{keV}$}. As a comparison, $^4$He production rate for the standard BBN is also presented (dotted blue line).}
\end{figure}

For temperatures of the order of $T\sim 80\;\rm{keV}$, this number is a very small quantity ($\sim 10^{-7}$) but a few neutrons nevertheless combine with ambient protons to form deuterium, which is then incorporated into helium nuclei. As the weak interactions are still efficient, some protons inverse-beta-decay to neutrons, restoring the equilibrium value. Given the timescale of 40 years, this very slow process ends up in an effective production of $^4$He nuclei. Fig. (\ref{dhedt}) present the production rate of $^4$He in both the Dirac-Milne universe (dashed line) and the standard cosmology (dotted line). We can discern the sharp decline in the production rate when the weak interactions decouple, leading to the nearly total disappearance of neutrons below $T \sim 80\;\rm{keV}$. As a simple argument illustrating how this mechanism leads to the right amount of $^4$He, we  note that the product of the $^4$He production rate with the Hubble time at production  is roughly the same in the two models. The production rates of $^4\rm{He}$ were obtained by calculating the time derivative of the $^4$He abundances.

To compute the primordial abundances of the different light elements, we solved the non-linear systems of first-order differential equations describing the network of the nuclear reactions
\be
\frac{dY_i}{dt}=\sum_rf^r_{kl}Y_kY_l-f^r_{ij}Y_iY_j, \quad i=1, N_{\rm{isot}}, \quad r=1, N_{\rm{reac}},
\ee
where $Y_i$ is the abundance of nuclide $i$, $N_{\rm{isot}}$  the number of nuclides , $N_{\rm{reac}}$ the numbers of nuclear reactions included in the network, $f^r_{ij}$ the rate of reaction $i+j\rightarrow k+l$, and $f^r_{kl}$ the rate of reaction $k+l\rightarrow i+j$.

To ensure accurate results, it is necessary to include a large number (over a hundred) of reactions.  As the time interval during which BBN takes place is very long, some slow reactions that are ineffective during SBBN must be integrated into the nuclear reaction network for the Dirac-Milne universe \citep{Simmering98}. Compared to previous studies of light element production in linear cosmology \citep{Kaplinghat00}, we used the nuclear reaction rates provided by \citet{nacre, CHJT2000, Tang03, Descouvemont2004}, as well as the usual rates of \citet{Wagoner69,Dicus82,CF88, FK90, Rau94}, and \citet{Chen1999}.

\begin{figure}
\includegraphics[width=\columnwidth]{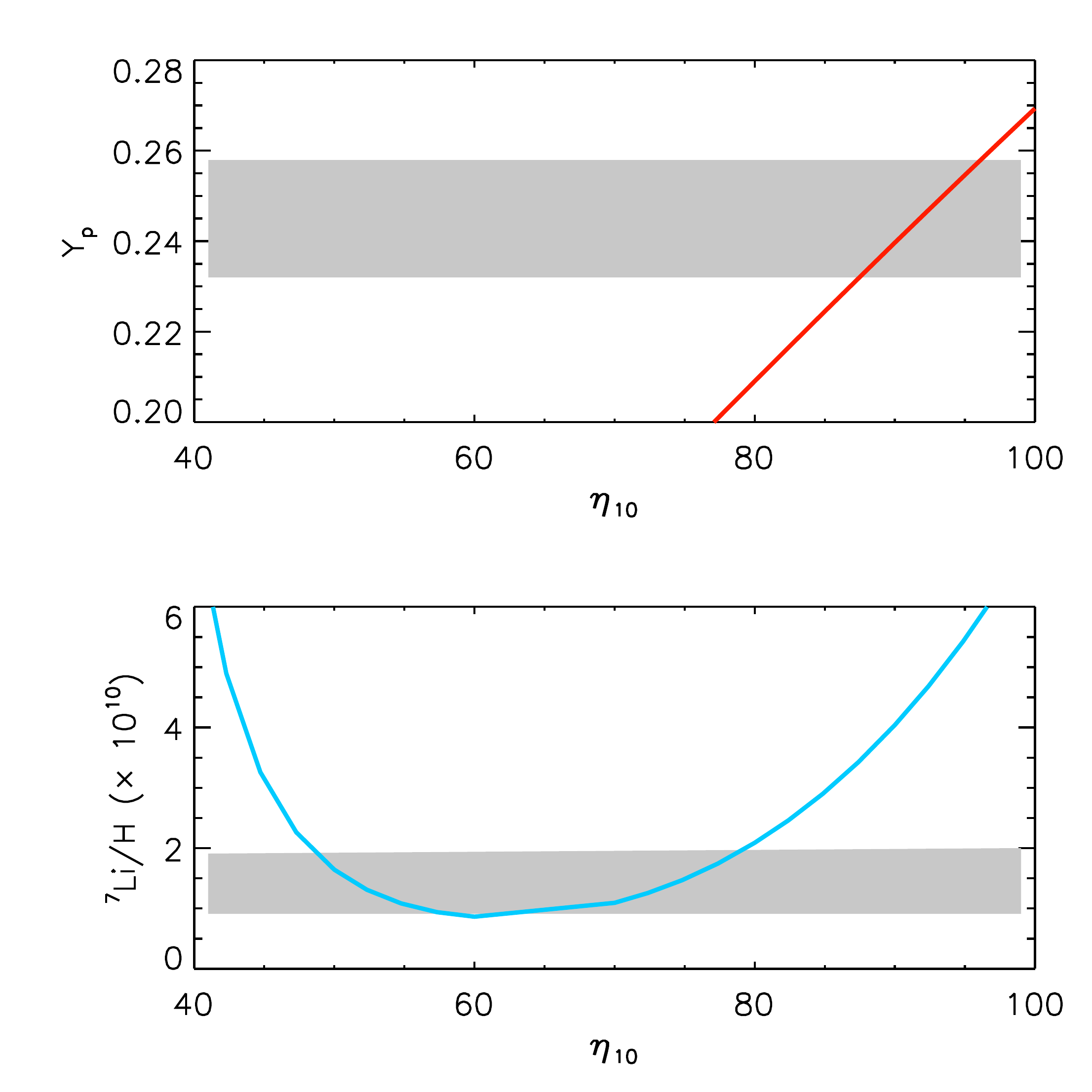}
\caption{\label{bbneta}$^4$He (top) and $^7$Li (bottom) theoretical abundances for the Dirac-Milne universe as a function of the baryon density. Here $\eta_{10}=10^{10} \eta$. Shaded areas correspond to the observational range. }
\end{figure}

The precise theoretical predictions for the primordial abundances strongly depend -- as in the standard model -- on the baryonic density. The final primordial abundances of $^4$He and $^7$Li as a function of the baryonic density, characterized by the ratio $\eta$ of the number of baryons to the number of photons ($\eta=n_b/n_\gamma$), are presented in Fig. (\ref{bbneta}). The shaded areas correspond to observational constraints \citep{Olive2004, Ryan2000} on the corresponding primordial abundances. The values of $\eta$ such that ensure that the theoretical value of $Y_p$ is compatible with the observations lie in the range
\be
8.8\times 10^{-9} \leq \eta\leq  9.6 \times 10^{-9}.
\ee

 As can be seen in Fig. \ref{bbneta} there is, as in the standard model,  no value of $\eta$ that permits the compatibility of both $^4$He and  $^7$Li with the model predictions. The lowest possible value of $\eta$ yields a lithium abundance of
\begin{equation}
\frac{^7\rm{Li} }{\rm{H}}= 3.45 \times 10^{-10}.
\end{equation} 
Although this value is somewhat larger than the one inferred by observations, we note that it is smaller than the predicted value in standard cosmology calculations, namely \mbox{$^7$Li/H$=(5.24_{-0.67}^{+0.71})\times 10^{-10}$} \citep{Cyburt08}. The Dirac-Milne universe clearly does not solve the $^7$Li problem, but nevertheless alleviates it. 

The value of the baryonic density inferred by the $^4$He observational constraint is almost 15 times higher than the value usually admitted in the framework of standard cosmology. This high baryonic density is an important feature of the Dirac-Milne universe as it suppresses the need for non-baryonic dark matter. Estimates of the dynamical mass in various structures all indicate a matter density in the Universe that is much higher than the baryonic density deduced by the BBN in the standard cosmology, which constitutes a strong motivation to postulate the existence of non-baryonic massive particles. In the Dirac-Milne universe, however, the BBN predicts a baryonic density comparable to the matter density estimated by different techniques. There is therefore no compulsory need for non-baryonic dark matter in the Dirac-Milne universe.

We note, however, that this high value of the baryonic density aggravates the so-called missing baryon problem \citep{Fukugita2004}. In the standard cosmology, this problem is that only approximately half the baryons predicted by SBBN are observationally detected. The other half is currently believed to reside in the warm-hot intergalactic medium (WHIM) \citep{Bregman07}. In the context of the Dirac-Milne universe, this WHIM would have to be the major source of baryons, although this possibility remains to be investigated. We note that attempts to explain the  dynamical behavior of galaxies consisting primarily of cold molecular (baryonic) gas instead of non-baryonic dark matter have been performed in the past \citep{Pfenniger94} and these scenarios would need to be revisited in the context of the Dirac-Milne model.

\begin{figure}
\includegraphics[width=\columnwidth]{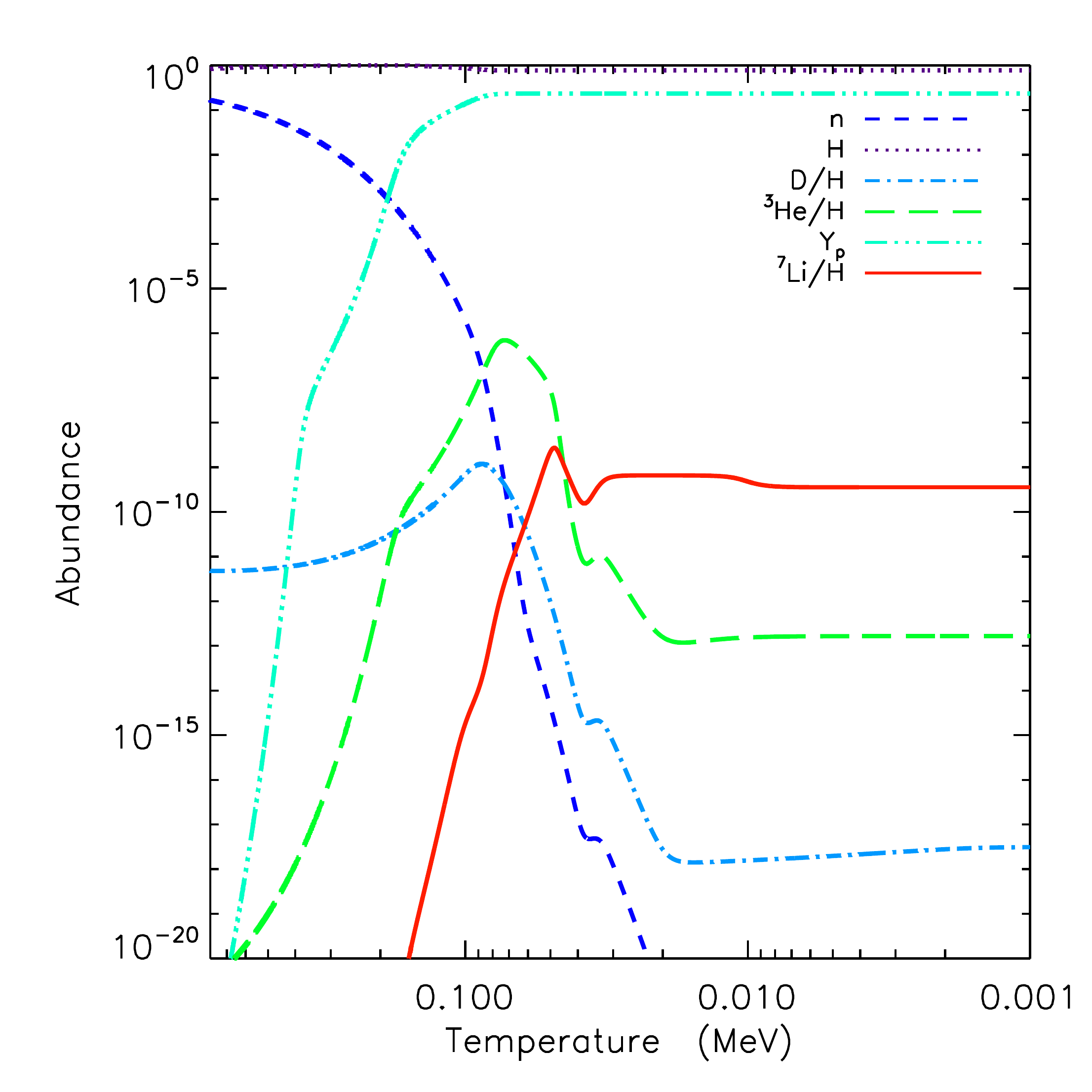}
\caption{\label{mbbn}Abundances of light elements obtained in the Milne universe with a baryon-to-photon ratio $\eta=8.8\times 10^{-9}$. Both $^4$He and $^7$Li are produced at observationally compatible levels (see Fig. \ref{bbneta}), but D and $^3$He are almost totally  destroyed by the very slow thermal nucleosynthesis.}
\end{figure}

Figure \ref{mbbn} presents the evolution of the primordial abundances for the light elements as a function of temperature for a baryonic density of \mbox{$\eta=8.8 \times 10^{-9}$}. It can be seen that D and $^3$He are almost totally destroyed during the stage of thermal production of $^4$He. This result was found to have severe consequences for linear cosmologies \citep{Kaplinghat00}. Although this is indeed the case for a regular linear cosmology without antimatter domains, the presence of distinct domains of matter and antimatter in the Dirac-Milne cosmology provides a natural scenario for the production of D and  $^3$He during a second stage of nucleosynthesis.

\subsection{Secondary production of D and $^3$He}

Studies of inhomogeneous big-bang nucleosynthesis as well as primordial nucleosynthesis in the presence of both  matter and antimatter domains have been conducted since the late 70's \citep{Combes75, Aly1978, Witten1984, Alcock1987, Applegate87, Kurki00b, Rehm01}. It has been shown that matter-antimatter annihilations at the frontiers of the domains can lead to the production of D, $^3$He, and T (later decaying as $^3$He) mainly through two channels of production: nucleodisruption ($\bar{p}^4\rm{He}$  and $^4\rm{He}\bar{^4\rm{He}} $ reactions) and photodisintegration of $^4$He nuclei. These studies provide us with the necessary material to compute the amount of deuterium produced by these various mechanisms.

Our purpose in what follows is to demonstrate the possibility of deuterium secondary production. We therefore consider that the emulsion has a static behavior, in the sense that its comoving size is assumed to remain constant. More precise studies investigating the dynamical behavior of the emulsion are beyond the scope of this first paper and will be treated in upcoming studies.

Annihilations at the frontiers of a domain are driven by the diffusion of nuclei towards the frontiers. The photodisintegration of $^4$He nuclei by energetic photons resulting from electromagnetic cascades induced by the annihilation photons and nucleodisruption are two possible processes that could produce deuterium.

\begin{figure}
\includegraphics[width=\columnwidth]{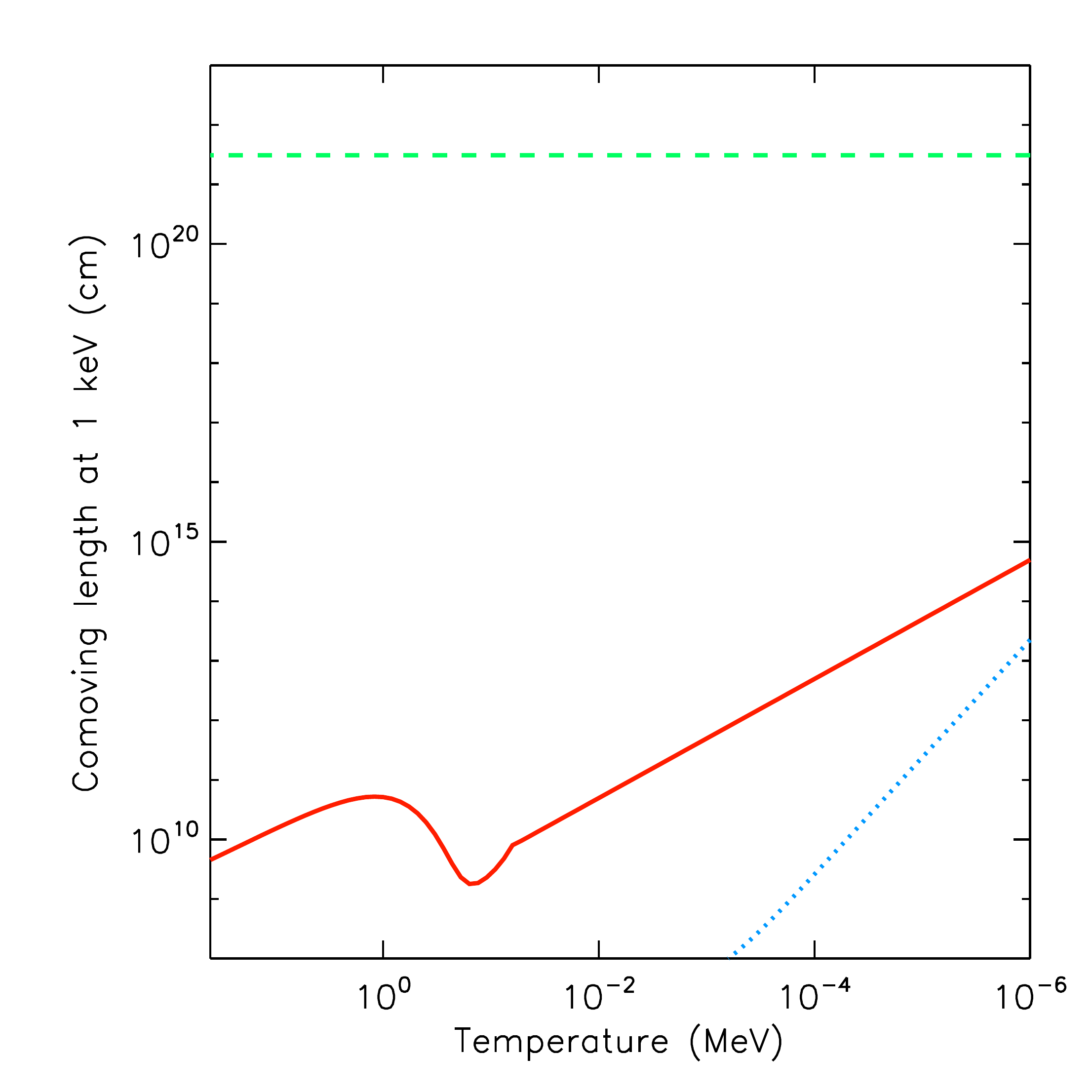}
\caption{\label{diff_length}   Comoving diffusion length (solid red line). The blue dashed line is the comoving thermalizing length of deuterium. Comoving Hubble distance is also shown in green.  }
\end{figure}

The main quantity to consider is the diffusion length. It represents the average distance  over which a (anti-)nucleus can diffuse toward the frontier of the domain on a Hubble time. This length gives an absolute lower bound to the size of the domains, as any concentration of (anti)matter smaller than this diffusion length would be annihilated during a Hubble time. The diffusion length is given by \citet{Applegate87}
\be
L_{\rm{diff}}(T)=\sqrt{6D(T)t_H(T)},
\ee
where $D$ is the diffusion coefficient and $t_H(T)$ is the Hubble time at temperature T. 
Using the diffusion coefficients given in \citet{Jedamzik01} and \citet{Sihvola01}, we computed the comoving diffusion length represented in Fig. \ref{diff_length}.

We can distinguish three regimes of diffusion. The first regime for $T\geq 1\;\rm{MeV}$ is regulated by neutron diffusion. Since neutrons are neutral particles, their electromagnetic interactions with other charged particles are very weak. As the temperature decreases, neutrons  disintegrate and diffusion is then maintained by protons,  which being charged have a much lower diffusion coefficient, which causes a dip in the diffusion length around 100 keV. As the density decreases because of the expansion, the diffusion length gradually increases until the temperature reaches $T\sim 50\;\rm{keV}$. At this temperature, the density is low enough for the mean distance between protons to be larger than the Debye length of electrons. Protons therefore do not behave as free particles but drag electrons along with them, ensuring charge neutrality. These electrons are themselves subject to Thomson drag \citep{PeeblesPPC} and thus limit the diffusion of protons \citep{Jedamzik01}. The effective diffusion coefficient of protons is then
\be
D_p^{\rm{eff}}=\frac{3T}{2\sigma_T\rho_\gamma},
\ee
where $\sigma_T$ is the Thomson cross-section and $\rho_\gamma$, the photon energy density.

\subsubsection{Annihilation rate}

In the most general case, the computation of the annihilation rate is a difficult task \citep{Aly_rate1978}. However, with our simplified approach based on diffusion, the estimation of the annihilation rate is rather straightforward. The annihilation rate is the number of annihilations per unit of time and surface. The quantity of matter (and antimatter) annihilated over a Hubble time {\bf per unit of surface} is $n_bL_{\rm{diff}}$, so that the annihilation rate is simply this quantity divided by the Hubble time
\be
\Psi=\frac{n_bL_{\rm{diff}}}{t_H}.
\ee
This simple expression is found to be in good agreement with the one derived in \citet{Cohen1998}. For simplicity,  and following previous studies of antimatter BBN \citep{Kurki00b, Rehm01}, we made the hypothesis that hydrodynamic turbulence, which could be produced by  energy release near the domain boundary, can be neglected.

\subsubsection{Production by photodisintegration of $^4$He nuclei}

The secondary production of light elements by $^4$He photodisintegration can occur in many scenarios and has been extensively discussed in the literature. In the framework of standard cosmology, this mechanism is known to produce D and $^3$He nuclei \citep{Ellis1992, Protheroe1995}. 

Proton-antiproton annihilations result in the production of neutral pions, which themselves decay to high-energy photons. Depending on the temperature of the background and their energy, these photons can create $e^+e^-$ pairs on CMB-photons. These newly created pairs can also interact with CMB photons and therefore lead to the creation of electromagnetic cascades \citep{Ellis1992, Protheroe1995}. These cascades stop when the energy of a photon becomes lower than the pair creation threshold $E_{\rm{pair}}=m_e^2/E_\gamma$, where $E_\gamma$ is the energy of a thermal photon. Owing to the high number of thermal photons, the threshold energy for pair creation is $E_{max}\sim  m^2_e/22T$ \citep{Ellis1992}. Photons with energy lower than $E_{\rm{max}}$ but higher than $E_c\sim m_e^2/80T$ also undergo elastic scattering on background photons \citep{Svensson1990}.
The resulting spectrum of cascaded photons can be parametrized by \citep{Ellis1992, Kurki00b}
\be
\left.\frac{dn_\gamma}{dE}\right|_{\rm{cas}}=\left\{
\begin{array}{ll}
A(E/E_c)^{-1.5},& E < E_c\\
A(E/E_c)^{-5},& E_c< E < E_{\rm{max}}\\
0,   &  E> E_{\rm{max}}
\end{array}\right. ,
\label{eq_spectre}
\ee
where $A= 3 E_0E_c^{-2}/ [7-(E_c/E_{\rm{max}})^3]$ is a normalization constant, and $E_0$ is the total energy injected.

Photodisintegration reactions ($^4$He($\gamma,\rm{p})^3$H, $^4$He($\gamma,$n)$^3$He, and $^4$He($\gamma,$np)D) require photons with energies higher than the respective threshold energies $Q_{^4\rm{He}(\gamma, np)\rm{D}}=26.07 \;\rm{MeV}$, $Q_{^4\rm{He}(\gamma, p)^3\rm{H}}=  19.81\;\rm{MeV}$, and $Q_{^4\rm{He}(\gamma, n)^3\rm{He}}=20.58\;\rm{MeV}$ \citep{Cyburt03}. The existence of these threshold energies implies that photodisintegration is a late process, as it becomes efficient only when $E_{\rm{max}}$ becomes higher than one of the previous threshold values. This happens at a temperature $T_{\rm{ph}}\sim 0.5\;\rm{keV}$.
\begin{figure}
\includegraphics[width=\columnwidth]{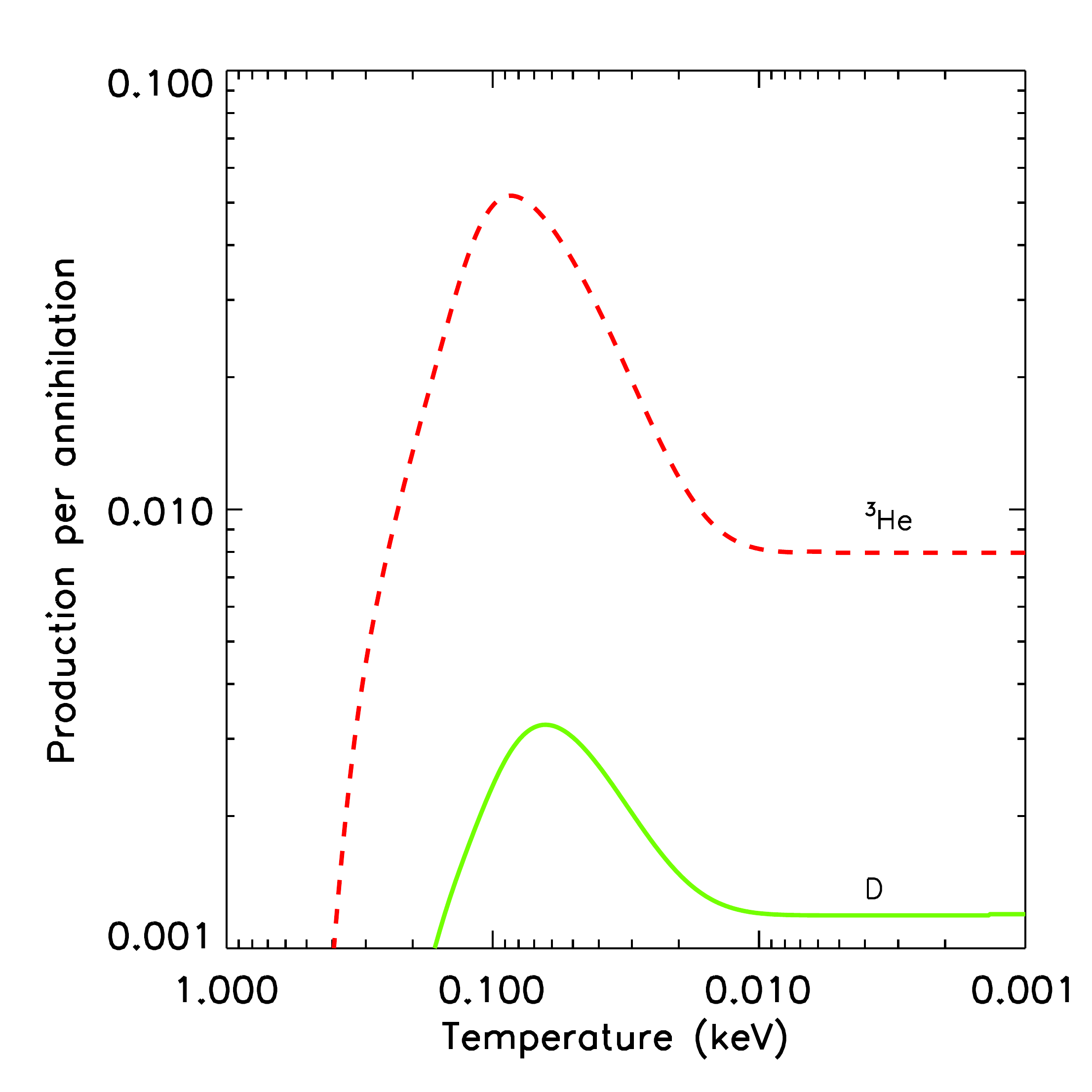}
\caption{\label{photo_p}   Quantity of D (solid green) and $^3$He (dashed red) nuclei produced by $^4$He nuclei photodisintegration for one $p\bar{p}$ annihilation. }
\end{figure}

High-energy photons responsible for $^4$He photodisintegration mainly interact with ambient electrons by means of Compton scattering or with ambient nuclei by means of the Bethe-Heitler process \citep{Jedamzik2006}. Taking this into account, the number of D nuclei produced per $p\bar{p}$ annihilation is given by
\be
N_{\rm{D}}=\int_{Q_{\rm{D}}}^{E_{\rm{max}}}dE_{\gamma} \frac{n_{\alpha} \sigma_{^4\rm{He}(\gamma,np)\rm{D}}(E_{\gamma})}  {f(E_\gamma)}\left.\frac{dn_{\gamma}}{dE_\gamma}\right|_{\rm{tot}}, 
\label{nd}
\ee
where $f(E_\gamma)=n_p\sigma_{BH}(\eg,1)+n_{\alpha}\sigma_{BH}(\eg,2)+k(\eg)n_e\sigma_{KN}(\eg)$, $\sigma_{BH}$ is the Bethe-Heitler process cross-section, $\sigma_{KN}$ is the Compton scattering cross-section in the Klein-Nishina regime \citep{RL79}, and $k(E_\gamma)\approx 1-{4/3}{[\ln(2E_\gamma/m_e)+1/2]^{-1}}$ is the mean fractional energy loss by Compton scattering \citep{Protheroe1995}. A similar formula exists for $^3$He. Figure \ref{photo_p} presents the results of this calculation.

When photodisintegration is most effective (around $T\sim 100\;\rm{eV}$), the mean free path of high energy photons relative to these processes is longer than the diffusion length, implying that D and $^3$He nuclei produced by photodisintegration will be able to survive and add to the overall light element production. More precisely, one can estimate the temperature (or equivalently the redshift) at which newly produced nuclei will not be annihilated later on. It is then necessary to assume that there exists a redshift $z_{\rm{end}}$ where matter and antimatter cease to annihilate. The hypothesis of gravitational repulsion between matter and antimatter leads to this {\it gravitational decoupling} but the exact mechanism, which would be analogous to the separation of electrons and holes in a gravitational field, still needs to be established. We emphasize that this decoupling is a necessary condition for the viability of  the Dirac-Milne model. A theoretical determination of $z_{\rm{end}}$  appears to be possible by analogy with the electron-hole system \citep{Tsidilkovski75}, but we assume here that it is a free parameter that will be later constrained by observations. 
We define $L_{\rm{diff}}^{z_{\rm{end}}}(z)$ the comoving value, calculated at a redshift $z$, of the diffusion length at redshift $z_{end}$ to be
\be
L_{\rm{diff}}^{z_{\rm{end}}}(z) =L_{\rm{diff}}(z_{\rm{end}})\left(\frac{1+z}{1+z_{\rm{end}}}\right).
\ee 
Any nuclei produced at a distance from the domain boundary smaller than this length will be annihilated by $z_{\rm{end}}$, whereas nuclei produced farther away will survive once annihilation stops. We can then determine the redshift $z_*$ below which the mean free path of high energy photons  is longer than $L_{\rm{diff}}^{z_{\rm{end}}}(z_*)$, where $z_*$ is then the redshift below which $D$ and $^3$He nuclei produced by photodisintegration will add up to the final abundances.

\begin{figure}
\includegraphics[width=\columnwidth]{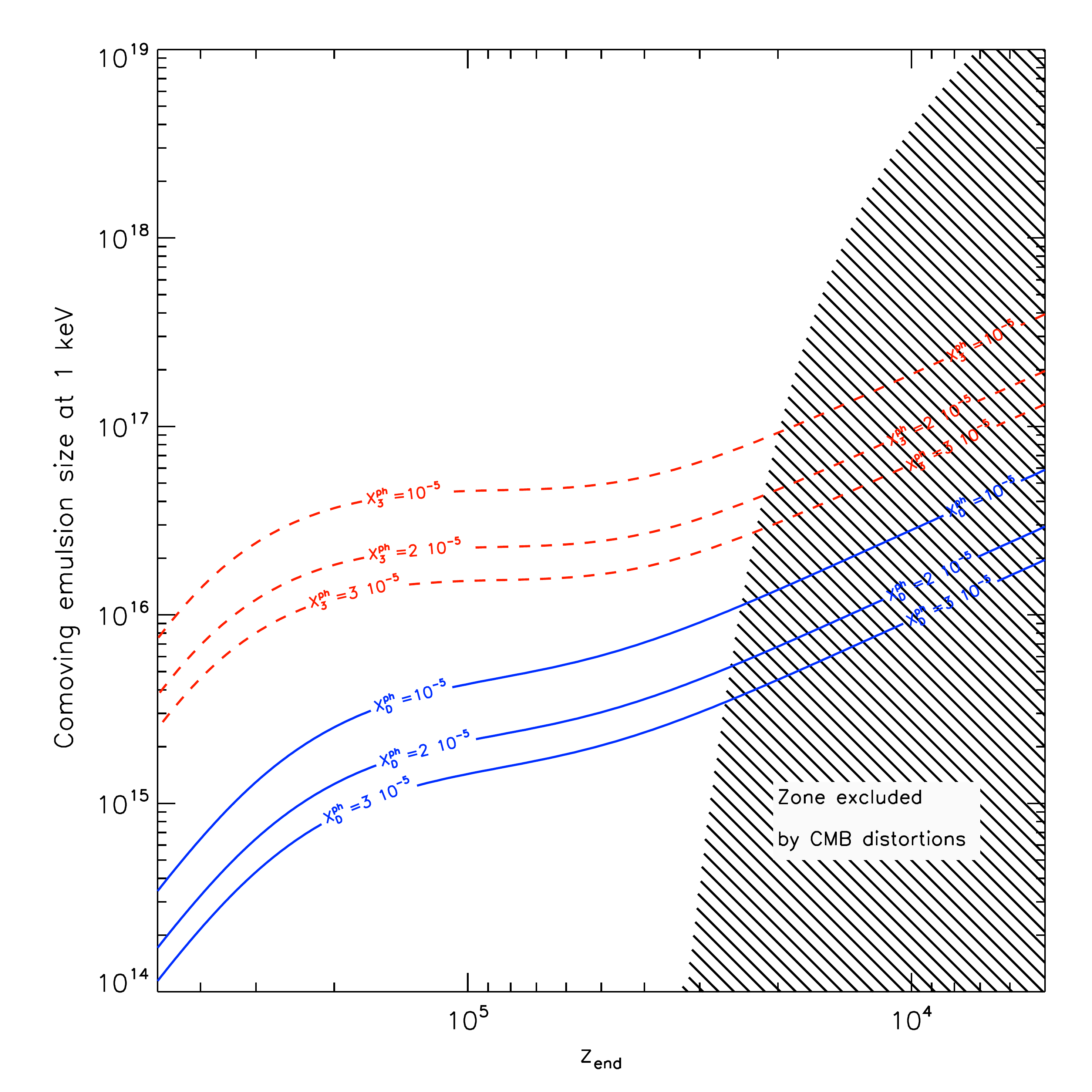}
\caption{\label{photo_pp} D and $^3$He final abundance as a function of the size of the emulsion and the redshift of gravitational decoupling. Hatched region is forbidden because of CMB distortions constraints.}
\end{figure}

Multiplying Eq. (\ref{nd}) by the annihilation rate and integrating between the times corresponding to $z_*$ and  $z_{\rm{end}}$, we get $N_{\rm{D}}^{\rm{ph}}$,  the number of D nuclei produced by photodisintegration of $^4$He nuclei
\be
N_{\rm{D}}^{\rm{ph}}=\frac{1}{H_0}\int^{z_{\rm{end}}}_{z_*}N_{\rm{D}}(z)\Psi(z)S(z)\frac{dz}{(1+z)^2}.
\label{ndph}
\ee
We can get rid of the surface term $S(z)$ by dividing eq. (\ref{ndph}) by the total number of baryons in the volume $V$, $n_b V$, which yields the final D abundance
\be
\left.\frac{\rm{D}}{\rm{H}}\right|_{\rm{ph}}=\left(\frac{T_0}{1\;\rm{keV}}\right) \frac{1}{L_{1\rm{keV}}}\int_{z_*}^{z_{\rm{end}}} L_{\rm{diff}}(z)N_{\rm{D}}dz.
\ee
This expression depends only on two variables, $z_{\rm{end}}$, the redshift of gravitational decoupling and $L_{1\rm{keV}}$, the comoving emulsion size at 1 keV. The values of the final D and $^3$He abundances as a function of these two parameters are presented on Fig. \ref{photo_pp}.

Electromagnetic cascades initiated by annihilation photons cause the injection of non-thermal energy into the background radiation. In the standard cosmology, these energy injections are tightly constrained, most notably by the FIRAS measurements \citep{Fixsen96}. However, in the Dirac-Milne universe, owing to the low value of the expansion rate, the radiative processes that can thermalize virtually any energy injection \citep{Danese77, Hu93} decouple at a lower redshift than in the standard case. The detailed calculation of the thermalization of CMB distortions will be presented elsewhere, but the exclusion contour in the $(z_{\rm{end}}-L_{1\rm{keV}})$ plane  is shown in Fig. \ref{photo_pp}.

To ensure that the final D abundance  agrees with observations ($\rm{D}/\rm{H}\sim 3\times 10^{-5}$, \citealp{Pettini08}) and imposing the gravitational decoupling as late as possible, we find that the comoving size of the domains  at 1 keV  has to be around
\be
L_{1\rm{keV}}\sim 10^{15}\;\rm{cm},
\ee
which corresponds to a comoving size of 7 kpc today. This estimate of the size of the emulsion may seem small compared, for instance, to the typical size of a galaxy, implying that there should be numerous matter and antimatter domains inside a single galaxy. However, we recall that this size is only constrained when gravitational decoupling occurs and that domains, which are in a non-linear mode of evolution immediately after photon matter-radiation decoupling, will have a dynamical behavior after recombination that effectively increases this size \citep{Dubinski93, Piran1997}. %Numerical simulations would probably help in having a deeper insight on that matter.

As suggested by Fig. \ref{photo_p}, the secondary production of light elements by photodisintegration of $^4\rm{He}$ leads to an overproduction of $^3\rm{He}$ relative to D. It is indeed usually assumed that D  can only be destroyed by stellar processes. On the other hand, $^3\rm{He}$ can be either produced or destroyed, but the ratio $\rm{D}/^3\rm{He}$ can only decrease \citep{Sigl1995}. Since this ratio is observed in our Galaxy to be  $\rm{D}/^3\rm{He\sim 1}$ but predicted to be $\rm{D}/^3\rm{He\sim 0.1}$ in the Dirac-Milne universe, the overproduction of $^3\rm{He}$ is a priori a strong constraint on the Dirac-Milne model. However, we note (see \citet{Jedamzik02} for a review) that D enhancement could occur at high redshift diminishing the efficacy of the $^3\rm{He}$ constraint.

\subsubsection{Production by nucleodisruption}

Nucleodisruption has been found to be a significant producer of D and $^3$He nuclei \citep{Sihvola01} within the standard evolution of the scale factor. However, with the slow evolution of the expansion rate in the Dirac-Milne universe and the hypothesis we made about the spatial repartition of the matter and antimatter domains, this situation changes.

Nuclei produced by nucleodisruption possess a kinetic energy ranging from a few MeV for nuclei to a few {\bf tens} of MeV for nucleons \citep{Balestra88}. These newly produced nuclei thermalize by Coulomb scattering on ambient protons and electrons. The thermalization length for D nuclei produced with an energy $E_0=10\;\rm{MeV}$ is presented in Fig. \ref{diff_length} (blue dotted line). This distance is always much smaller than the diffusion length, implying that any D nucleus produced by nucleodisruption will finally return towards the annihilation zone and be destroyed there. A possible way to produce a higher fraction of deuterium by nucleodisruption would be to consider small domains of (anti)matter within a larger domain  of antimatter (matter). This situation occurs continually in an emulsion, which suffers a redistribution of ``domains" when bridges in the emulsion disappear by annihilation. If the dimension of the larger domain is larger than the diffusion length, then an important fraction of the D and $^3$He produced by nucleodisruption could survive. However, precise calculations of this production require the knowledge of the statistical properties of the spatial distribution of domains, which strongly depends of the separation mechanism. This point should be investigated in future studies of nucleosynthesis in the Dirac-Milne universe.

\section{Type Ia supernov\ae\label{snia}}

In 1998 \citep{Riess98, Perlmutter99}, distance measurements for type Ia supernov\ae\ (SNe Ia) revealed that these objects are dimmer than expected if our Universe was correctly described by a decelerating Einstein-de Sitter model. The introduction of a cosmological constant $\Lambda$ in the field equations of general relativity, which is apt to produce an accelerating expansion, provided an impressive fit to the observational data. Today, SNe Ia are one of the most important cosmological tests and are considered as prime evidence of an acceleration of the expansion. We recall however that the strong evidence of a recent transition between a decelerating phase and an accelerating phase of expansion heavily relies on the prior hypothesis of spatial flatness. Without this hypothesis, the evidence is less clear \citep{Seikel2008}.

The Dirac-Milne universe has neither acceleration nor deceleration and is therefore equivalent to an open empty universe. In terms of the usual cosmological parameters, this universe corresponds to the combination
\be\left(\Omega_M=0, \Omega_\Lambda=0\right).
\ee

In this context, the luminosity distance in the Dirac-Milne universe follows the simple expression
\be
d_L(z)=\frac{c}{H_0}(1+z)\sinh[\ln(1+z)]. 
\ee

It is usually claimed that the empty universe, hence the Dirac-Milne universe,  is strongly inconsistent with the SNe Ia observations. We wish here to elaborate on this statement using the data of the first release of the SNLS collaboration \citep{Astier06}. The SNLS data consist of two distinct datasets. The high-redshift sample, from the SNLS, comprises 71 SNe Ia with redshifts between $0.2\leq z\leq 1.01$. The second sample is a low-redshift set, consisting of 44 SNe Ia taken from the literature with redshifts $z\leq 0.15$. These data come from different experiments and are therefore possibly subject to different sources of systematic errors. 

Following the  definition given in \citet{Astier06}, the distance modulus is
\be
\mu_B=m^*_B-M+\alpha(s-1)-\beta c, 
\ee
where $M$ is the absolute magnitude of SNe Ia, $\alpha $ and  $\beta$ are global parameters that link the stretch $s$ and the color $c$ to the distance modulus, and $m^*_B$ is the apparent magnitude of the supernova. 
It should be emphasized that, in contrast to the $\Lambda$CDM cosmology, there is no cosmological parameter dependence in the Dirac-Milne luminosity distance. The only degrees of freedom are the nuisance parameters, $M$, $\alpha$, and $\beta$.
Following the procedure described in \citet{Astier06}, we minimize the expression
\be
\chi^2=\sum\frac{(\mu_B-5\log_{10}(d_L/10\;\rm{pc}))^2}{\sigma^2(\mu_B)+\sigma_{\rm{int}}^2}.
\label{chi2}
\ee
Here, $\sigma(\mu_B)$ takes into account measurement errors in the apparent magnitude $m_B^*$, stretch, and color parameters derived by the analysis of light curves \citep{Guy2005}, $\sigma_{\rm{int}}$ is the so-called "intrinsic" dispersion, which is a parameter introduced to account for SNe Ia being astrophysical objects that naturally have some intrinsic dispersion in their absolute magnitude. However, the value of this parameter is unknown, and  in the fitting procedure, $\sigma_{\rm{int}}$ is adjusted to ensure that the reduced chi-squared is unity. 

\subsection{Analysis with only the  high-z sample}

We first performed our analysis on the high-$z$ sample without including any low-z SN Ia. Without this low-$z$ anchoring, the analysis does not permit us to discriminate between the $\Lambda$CDM and the Dirac-Milne universes. In this respect, we note that the three-year analysis of SNLS using their data alone \citep{Guy10} is consistent to a better than 68\% CL with the Dirac-Milne universe, while the Einstein-de Sitter (EdS) model is clearly excluded. The evidence of an expansion acceleration therefore relies on a comparison between low-$z$ and high-$z$ SNe Ia. We also present the results for the EdS model.

\begin{table}[htb]
\begin{center}
\caption{\label{tab:sn5}Values of the different parameters for the fit using only the 71 high-$z$ SNe Ia of the SNLS. Here, $\sigma_{\rm{int}}$ has been fixed to $\sigma_{\rm{int}}=0$. The number in parenthesis indicates the number of degrees of freedom. }
\begin{tabular}{cccc}
\hline
\hline
       Parameter           			& Dirac-Milne			& $\Lambda$CDM & EdS	\\
             \hline
$\Omega_M$ 	& $\cdots$		                  	 & $0.289    \pm 0.033 $	 & $1$ \\
%\hline
%$ M$		                 & $-19.24\pm 0.006$ 	& $-19.36\pm 0.02 $ & $-19.04\pm0.01$	\\
$ M$		                 & $-19.24$ 	& $-19.36 $ & $-19.04$	\\

%\hline
$\sigma_{\rm{int}}$	& 0	& 0 &0 \\
\hline
$\chi^2$ total			&553.64				&558.6 & 724.38 	\\
%\hline
$\chi^2$/dof   	& 8.14(68)  &  8.33 (67) & 10.65(68)\\
%\hline
\end{tabular}
\end{center}

\end{table}

\begin{figure*}
\includegraphics[width=17cm]{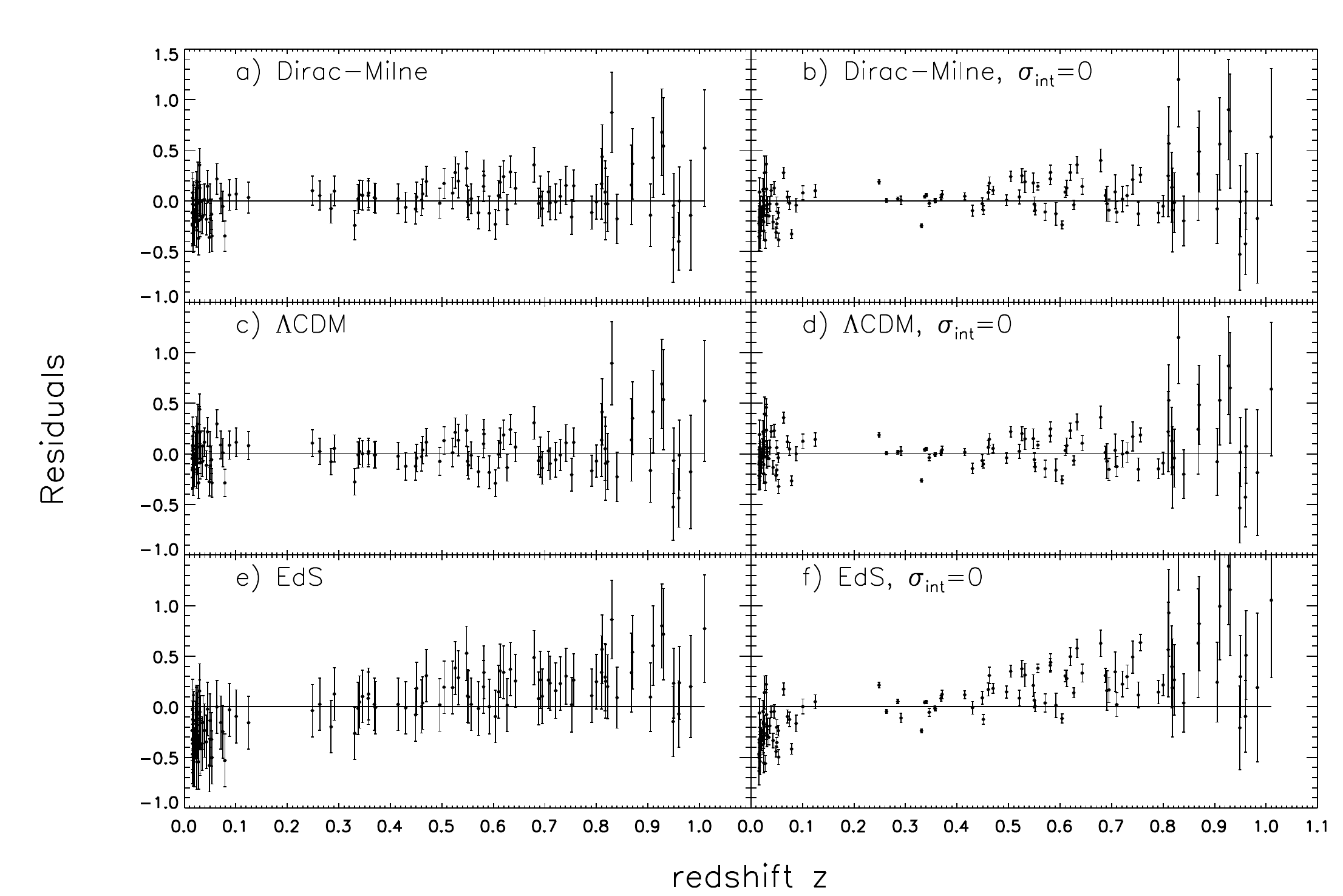}
\caption{\label{res_sn}Residuals of the Hubble diagram for the Dirac-Milne ((a) and (b)), flat $\Lambda$CDM ((c) and (d)), and Einstein de-Sitter ((e) and (f)) models. The left column represents the residuals obtained by the minimization of the $\chi^2$ defined by eq. (\ref{chi2}), in which the values of the intrinsic dispersion is adjusted so that $\chi^2/\rm{d.o.f.}=1$. In the right column, this intrinsic dispersion is fixed to 0.}
\end{figure*}

\begin{table}[htb]
\begin{center}
\caption{\label{tab:sn4}  Values of the different parameters for the fit using only the 71 high-$z$ SNe Ia of the SNLS. Here $\sigma_{\rm{int}}$ has been adjusted such that $\chi^2/\rm{d.o.f.}=1.$ The number in parenthesis indicates the number of degrees of freedom.}

\begin{tabular}{cccc}
\hline\hline
               Parameter           			& Dirac-Milne			& $\Lambda$CDM  &EdS	\\
                  \hline
$\Omega_M$ 	&  $\cdots$		                  	 & $0.25    \pm 0.08 $	 &1\\
%\hline
%$ M$		                 & $-19.18\pm 0.022$ 	& $-19.33\pm 0.07 $ 	&$-18.87\pm 0.03$\\
$ M$		                 & $-19.18$ 	& $-19.33 $ 	&$-18.87$\\
%\hline
$\sigma_{\rm{int}}$	& 0.1172	& 0.1173  & 0.155\\
\hline
total $\chi^2$ 			&68.01				&67.01&  67.97	\\
%\hline
$\chi^2$/dof  	& 1.00 (68)  &  1.00 (67) & 1.0 (68)\\
%\hline
\end{tabular}
\end{center}
\end{table}

The results of the analysis of the 71 SNe Ia are presented  in Tables \ref{tab:sn5} and \ref{tab:sn4}. In Table \ref{tab:sn5}, the intrinsic dispersion is fixed to a null value. The analysis is therefore performed using only the measurement errors, thereby giving a stronger weight to SNe Ia with redshifts $0.2 \leq z\leq 0.4$, which have smaller errors. 
The total and reduced $\chi^2$ of the EdS model are much larger than those of the $\Lambda$CDM and Dirac-Milne models, which are in turn similar.
In Table \ref{tab:sn4}, the intrinsic dispersion is determined by assuming that the reduced chi-squared is unity. The value of the intrinsic dispersion required to renormalize the reduced $\chi^2$ of the EdS model to unity is here again much larger than those required for the other two models. The conclusion of this first analysis is twofold. First, we confirm that the decelerating Einstein-de Sitter model is extremely unlikely, which should come as no surprise. Second, the Dirac-Milne and the flat $\Lambda$CDM models are almost identical, Dirac-Milne being in even closer agreement with the data than the flat $\Lambda$CDM model. Our analysis stresses that all SNe Ia analysis depend strongly on the use of low-z data to anchor the Hubble diagram.

\subsection{Analysis on the full data set}

Similarly, we proceeded with our analysis of the full data sample used by SNLS in its one-year analysis, i.e. using an heterogeneous sample of low-$z$ SNe Ia. Our results are given in \mbox{Tables \ref{tab:sn6} and \ref{tab:sn7}}. In this case, as expected, the flat $\Lambda$CDM provides a closer fit to the data than the Dirac-Milne universe. 

\begin{table}[htb]
\begin{center}
\caption{\label{tab:sn6} Values of the different parameters for the fit using the complete data set of 115 SNe Ia. Here, $\sigma_{\rm{int}}$ has been fixed to $\sigma_{\rm{int}}=0$. The number in parenthesis indicates the number of degrees of freedom. }

\begin{tabular}{cccc}
\hline\hline
          Parameter        			& Dirac-Milne			& $\Lambda$CDM & EdS	\\
                  \hline
$\Omega_M$ 	&  $\cdots$			                  	 & $0.258    \pm 0.019 $	&1 \\
%\hline
%$ M$		                 & $-19.258\pm 0.006$ 	& $-19.38\pm 0.011 $ &$-19.01\pm0.008$	\\
$ M$		                 & $-19.258$ 	& $-19.38$ &$-19.01$	\\
%\hline
$\sigma_{\rm{int}}$	& 0			& 0		&0 \\
\hline
$\chi^2$ total			&877.93				&809.75  & 	1507.83\\
%\hline
$\chi^2$/dof  	& 7.83 (112)  &  7.295 (111) & 13.46 (112)\\
%\hline
\end{tabular}
\end{center}
\end{table}

\begin{table}[htb]
\begin{center}
\caption{\label{tab:sn7}    Values of the different parameters for the fit using  the  complete data set of 115 SNe Ia. Here $\sigma_{\rm{int}}$ has been adjusted such that $\chi^2/\rm{d.o.f.}=1.$The number in parenthesis indicates the number of degrees of freedom.   }

\begin{tabular}{cccc}
\hline\hline

           Parameter       			& Dirac-Milne			& $\Lambda$CDM 	&EdS	\\
                  \hline
                  $\Omega_M$ 		&  $\cdots$ 		    	 & $0.250    \pm 0.036 $	&1\\
%\hline
%$ M$		                 & $-19.217\pm 0.020$ 	& $-19.331\pm 0.029 $& $-18.98\pm0.035$	\\
$ M$		                 & $-19.217$ 	& $-19.331 $& $-18.98$	\\

%\hline
$\sigma_{\rm{int}}$	& 0.1432				& 0.1289		& 0.258		 \\
\hline
$\chi^2$ total			&112.05				&110.96  & 	111.95\\
$\chi^2$/dof		&1.00	(112)			& 0.99(111)		& 0.99	(112)	\\
%\hline
\end{tabular}
\end{center}
\end{table}

In Fig. \ref{res_sn}, we present the residuals of the Hubble diagram for the Dirac-Milne, the flat $\Lambda$CDM, and Einstein-de Sitter models. The left-hand column represents the residuals when the value of the intrinsic dispersion was adjusted to normalize the  $\chi^2$ to 1 per degree of freedom. In the right-hand column,  we present the residuals obtained when the intrinsic dispersion parameter is fixed to the value $\sigma_{\rm{int}}=0$. Setting $\sigma_{\rm{int}}$ to 0 enables one to consider the ``real"  measurement errors. It appears that for SNe with redshifts in the range $0.2 \leq z\leq0.8$ these errors are smaller than the intrinsic dispersion. The use of such an ad-hoc parameter in the analysis  may degrade the quality of the data in the redshift interval $0.2-0.4$, where the quality of the observations is highest. 
We note that such an analysis with statistical errors only was previously performed by  the SNLS three-year analysis \citep{Guy10}.

Panels e) and f) present the residuals for the Einstein-de Sitter model. Even though the reduced $\chi^2$ has been constrained to unity in panel e), the characteristic slope in the residuals illustrates the non-conformity of the  Einstein-de Sitter model to the data. The difference between the Dirac-Milne and the $\Lambda$CDM models is however almost negligible, and can hardly be seen by simply examining the residuals.

We note that it is only the use of nearby SNe Ia that enables us to distinguish between the Dirac-Milne and the $\Lambda$CDM models, the latter then being the most likely model at a confidence level of more than 3$\sigma$ as previously announced \citep{Astier06, Kowalski2008}.

It appears that sources of previously unaccounted for systematic errors \citep{Kelly2010} are present in the nearby sample data set. To investigate this, we determined  the constant offset to the apparent magnitude of nearby SNe Ia required for the chi-squared for the Dirac-Milne and the flat  $\Lambda$CDM to become equal. We found that an offset of $\delta m^*_B=0.06 \;\rm{mag}$ is sufficient to ensure that the two models are equally probable. This value should be compared to the budget of systematics errors estimated in recent studies: $\Delta M=0.04\;\rm{mag}$ \citep{Kowalski2008}. Therefore, a relatively mild systematic error of 1.5 $\sigma$ for nearby SNe Ia would lead us to favor the Dirac-Milne universe over the conventional $\Lambda$CDM cosmology in the SNe Ia analysis.

\section{Other tests\label{CMB}}

A major result of CMB experiments has been the precise measurement of  the position of the first acoustic peak on the degree scale, which seems to imply that the spatial curvature is nearly zero \citep{WMAP7K}. In the open spatial geometry of  the Dirac-Milne universe, this position would naively be expected at a much smaller angle. The ratio of the angular distances in the two models taken at redshift  $z \sim 1100$, which corresponds to the surface of last scattering surface, is

\be
\frac{d_A^{\rm{Milne}}(z)}{d_A^{\Lambda CDM}(z)}\stackrel{z= 1100}{\sharp}169.
\ee
The value of this ratio implies that an astrophysical object at redshift $z=1100$ is seen under an angle 169 times smaller in the Dirac-Milne universe than in the $\Lambda$CDM cosmology. 

 The angular position of the first peak is defined by the angle under which the sound horizon is seen at recombination
 \be
 \theta=\frac{\chi_s(z_*)}{d_A(z_*)},
 \ee
 where $\chi_s(z_*)$ is the sound horizon, $d_A(z_*)$  the angular distance, and $z_*$ the redshift of the last scattering surface. It is of interest to consider the equivalent multipole $\ell_a\sim \pi/\theta$. The sound horizon is defined as the distance  that acoustic waves can travel in the primordial plasma. Taking into account the universe expansion, this distance reads
 %Neglecting the small variation in the speed of sound, $c_s$ induced by the presence of baryons, it reads:
\be
\chi_s=\int_0^tc_s\frac{dt^\prime}{a(t^\prime)}, 
\ee
where the speed of sound $c_s=c/\sqrt{3(1+R)}$, $R$ being a corrective factor caused by the presence of baryons \citep{Hu1995}. Its value can be related to the baryon to photon ratio $\eta$ by \mbox{$R\approx 1.1\times 10^{12} \eta/(1+z)$}.

The definition of the lower bounds of the integral requires some care as we have shown previously that this integral diverges near the initial singularity. However, the mechanisms of sound generation in the Dirac-Milne and $\Lambda$CDM universes differ radically. In contrast to the generation of sound waves in the standard model where inhomogeneities are produced at the epoch of inflation, it seems reasonable to consider that sound waves in the Dirac-Milne universe are probably produced by annihilation at the matter-antimatter frontiers. It is therefore natural to consider for this value the epoch of QGP transition around $T\sim 170 \;\rm{MeV}$. In the absence of inflation, the universe is indeed very homogeneous before this event, and QGP transition seems naturally to be the scale of interest in the Dirac-Milne cosmology. There is at present a rather general consensus that the QCD transition is of neither first nor second order at null chemical potential, but is probably an analytic crossover (see e.g. \citet{Aoki06, Bazazov09, Endrodi11}, and references therein). Under this hypothesis, it would be extremely difficult, if not impossible, to understand how significant amounts of matter and antimatter could have survived annihilation in the very slow evolution of the Dirac-Milne universe. However, this present belief is based on a QCD calculation, whereas the situation of the primordial universe is more complicated, including additional light leptons (electrons, positrons, neutrinos, and antineutrinos). In addition, present calculations use light quark masses for the u and d quarks that are significantly higher than the actual masses of these quarks. Therefore, the observation of large matter-antimatter domains would be an extremely useful indication that, in contrast to present expectations, there is a sharp transition allowing the survival of significant regions of antimatter at the QCD transition. It should also be noted that some authors are clearly considering the possibility that baryogenesis occurs at the QCD transition or at a $O$(100 MeV) temperature (see for example \citet{Dolgov92} for a review).

 Acoustic waves then propagate in the plasma as long as matter and antimatter are in contact, {\it i.e.}  until the gravitational decoupling, estimated in the previous section at $z_{end} \approx 3\times 10^{4}$. With these values, the comoving sound horizon is found to be
\be
\chi_s\sim 42\;\rm{Gpc}.
\ee 
%After this epoch, the sound horizon simply follows the expansion of the universe.
The expression of the angular position of the first acoustic peak then follows

\beqn
\ell_a&\sim &\pi \frac{d_A}{\chi_s(z_*)},
  \label{acoustic}
\eeqn

It can be shown that the redshift of the last scattering surface in the Dirac-Milne universe is a few percent lower than in the standard cosmology, here again due to the late decoupling of the radiative processes leading to recombination. We found that $z_*\sim 1040$.

%\begin{equation}
%l_A=\frac{180}{\pi}\left(\int^{z_{i}}_{z_{\rm{stop}}}\frac{1}{\sqrt{3}}\frac{dz}{1+z} \right)\times \left(\frac{1}{\sinh \ln(1+z_*)}\right),
%\label{peak}
%\end{equation}

Calculating the multipole of the acoustic scale using expression (\ref{acoustic}), we obtain $\ell_a\sim 160$. The standard value of this quantity is $\ell_a\sim 300$ \citep{WMAP1}. Instead of a discrepancy of a factor $\approx 169$ , there is an almost exact compensation between the larger geometrical term, induced by the open geometry of the Dirac-Milne universe, and the larger sound horizon, caused by the slow evolution of the expansion rate before recombination. Taking into account the numerous approximations in the model, this remarkable coincidence is quite unexpected and represents a fascinating motivation to study the Dirac-Milne universe in more detail.

However, the sound horizon scale is also imprinted in the large-scale structure power spectrum under the form of small oscillations \citep{Eisenstein98} called baryonic acoustic oscillations (BAOs).   These oscillations are expected, at least in  a first approximation, on the same scale as the sound horizon. As discussed above, the sound horizon is much larger in the Dirac-Milne universe than in the standard cosmology. Admitting that there are BAOs in the Dirac-Milne universe, they should be expected on a scale much larger than that of the standard cosmology. The claimed detection of these BAOs on the expected scale (within the standard cosmology) of \mbox{$\sim 150\;\rm{Mpc}$} \citep{Eisenstein05}, presently detected at the $\sim 3 \sigma$ level, if confirmed by ongoing experiments, would therefore provide a strong constraint on the Dirac-Milne universe.

\section{Conclusion}

%Motivated by the puzzling situation in cosmology, 
Since the standard $\Lambda$CDM model is in good agreement with observations but rather poorly theoretically motivated, we have studied here an alternative cosmological model, the Dirac-Milne universe. Inspired by the work of Dirac, Kerr, and Carter, this model restores the symmetry between matter and antimatter. Relying on the symmetries of the Kerr-Newman solutions in general relativity, it makes the hypothesis that particles and antiparticles behave similarly to quasiparticles such as electrons and holes in a semiconductor, and that antimatter has a negative active gravitational mass. A fundamental characteristic of this Universe is the linear evolution of its scale factor, which solves in an elegant way both the problems of the horizon and the age of the universe.

For primordial nucleosynthesis, we have found that the Dirac-Milne universe is able  to produce $^4$He at an adequate level, while producing $^7$Li nuclei in proportions admittedly a factor three higher than the observed values but with a smaller disagreement between observations and predictions than the standard cosmology. We have also shown that surface annihilations at the frontiers of matter and antimatter naturally lead to the production of D nuclei, in amounts that are proportional to the inverse of the characteristic size of the emulsion.  The main focus of this study has been the production of D, although the assumption that the emulsion has a fixed characteristic size is clearly an approximation. Relaxing this assumption will allow the possibility of the total annihilation of small patches of antimatter inside larger regions of matter. This might lead to a net production of D by nucleodisruption, which has a higher D/$^3$He production ratio \citep{Balestra88}, hence alleviate the constraints made by the overproduction of  $^3$He nuclei .

The Dirac-Milne universe does not undergo an accelerated expansion and therefore seems to conflict with the usual interpretation of type Ia distance measurements. We have shown that the Dirac-Milne universe can nevertheless be reconciled with type Ia supernov\ae\  observations if we take into account a reasonable size of systematic errors in the low-z supernov\ae\ subset.  Finally, and perhaps the most surprising result of this unusual cosmology is that the acoustic scale naturally emerges at the degree scale, despite the open geometry.

These first results are encouraging but some issues have not been addressed in this study. Amongst these is how  structure formation occurs in the presence of separate domains of positive and negative mass. The use of numerical simulations will most probably be a necessity as the usual linear approximation does not hold. Immediately after matter-radiation decoupling, density contrast is indeed on the order of unity for any distribution of matter with positive mass and antimatter with negative mass.

 \section*{Acknowledgments}
 
 It is a pleasure to acknowledge fruitful discussions with J. Andrea, E. Armengaud, B. Carter, N. Fourmanoit, J. Fric, K. Jedamzik, T. Jolic\oe ur, E. Keihanen, R. Pain, J. Rich, and the members of the SNLS collaboration. We express our special thanks to A. Coc for allowing us to adapt his nucleosynthesis code. We also thank an anonymous referee for his/her questions and comments. Needless to say, these people are not responsible for the errors present in this admittedly provocative but hopefully interesting paper.

%\end{acknowledgments}

%\bibliography{/Users/gabrielc/Documents/Biblio_references/biblio2}% Produces the bibliography via BibTeX.
\bibliographystyle{aa}
\bibliography{/Users/benoitl/Documents/Post_doc/Papers/biblio2}
%\bibliography{/Users/gabrielc/Documents/Articles/Milne_Universe_AetA/biblio2}

%\begin{thebibliography}{99}

%\end{thebibliography}

\end{document}